# Size dependence of the effective magnetic anisotropy in Co, Ni, Fe, and magnetite nanoparticles: Testing the core-shell-surface layer (CSSL) model

Sobhit Singh and M. S. Seehra*

*Department of Physics & Astronomy, West Virginia University, Morgantown, WV-26506, USA*

*Corresponding author: Mohindar.Seehra@mail.wvu.edu

## Abstract

The stability of the stored information in recording media depends on the anisotropy energy $E_a = K_{eff}V$ of nanoparticles (NPs) of volume $V$ or diameter $D$. Therefore, the knowledge of how the effective magnetic anisotropy $K_{eff}$ varies with $D$ for a given system is important for technological applications. In a recent paper [Appl. Phys. Lett. 110 (22), 222409 (2017)], the variation of $K_{eff}$ versus $D$ in NPs of maghemite (γ-Fe$_2$O$_3$) was best described by the Eq.: $K_{eff} = K_b + (6K_S/D) + K_{sh}\{[1-(2d/D)]^{-3} -1\}$, where $K_b$, $K_S$, and $K_{sh}$ are the anisotropy constants of spins in the core, surface layer, and a shell of thickness $d$, respectively. This core-shell-surface layer (CSSL) model is an extension of the often used core-surface layer (CSL) model described by $K_{eff} = K_b + (6K_S/D)$ [Phys. Rev. Lett. 72, 282 (1994)]. The additional term involving $K_{sh}$ was found to be necessary to fit the data for smaller NPs of γ-Fe$_2$O$_3$ with $D$ < 5 nm. Here we check the validity of the CSSL model for metallic magnetic NPs of Co, Ni, Fe and magnetite using $K_{eff}$ vs. $D$ data from published literature. Care was taken in selecting data only for those NPs for which the effects of interparticle interactions has been taken into account in determining $K_{eff}$. The importance of the new CSSL model is that it describes well the $K_{eff}$ vs. $D$ variation for all particles sizes whereas the core-surface layer model often fails for smaller particles with the notable exception of Fe NPs. The verification of the CSSL model for metallic NPs of Co, Ni, and magnetite along with NPs of NiO and γ-Fe$_2$O$_3$ validates its general applicability.

1. **Introduction:** The increasing demand of magnetic nanoparticles (NPs) for applications in compact magnetic storage media, catalysis, ferrofluids, sensors, magnetic drug delivery, and biomedicine have secured a unique place for nanoparticle research in the scientific community. [1, 2, 3, 4, 5] A particularly interesting feature of magnetic NPs is their size-dependent magnetic properties, both due to finite size effects and the increasing role of surface spins with decreasing particle size. With decreasing particle size ($D$), the concentration of unsaturated surface spins increases as $1/D$ causing reduction in the net magnetization and enhancement in the effective magnetic anisotropy ($K_{eff}$) of NPs. Details of the size-dependence of $K_{eff}$ is of primary interest for applications in magnetic data storage technology since the stability of the stored information in recording media depends on the anisotropy energy $E_a = K_{eff}V$ of NPs of volume ($V$). Large anisotropy energy is desired to keep the stored information robust against the thermal activation of spins. Generally, a ratio $K_{eff}V/k_BT > 40$ is required for reliable storage of data for ~10 years, $k_BT$ being the average thermal energy [1].

The spins on the surface of the NPs experience a different anisotropy compared to those in the bulk (core) of the NPs due to the broken exchange bonds and reduced crystalline symmetry of the surface. Taking this fact into consideration, Bodker *et al.* proposed a core-surface layer (CSL) model to describe the linear trend of $K_{eff}$ *versus* $1/D$ data for Fe NPs and to separate the contributions of the surface and bulk spins in the total effective anisotropy energy of magnetic NPs [6]. This CSL model was successful to describe the $K_{eff}$ *versus* $1/D$ data for a large range of magnetic NP systems; However, deviations from this model have been reported for ultra-fine magnetic NPs. [7, 8, 9, 10, 11, 12, 13] The main reason for the limitation of the CSL model for ultra-fine magnetic NPs is that the model does not account for the partially ordered spins in the shell layer. Recent experimental, theoretical and computational studies have shown that the surface spin disorder in NPs is not localized at the surface layer only, but it tends to gradually propagate towards the core forming a shell of finite thickness $d$ [14-18] making the ordering of spins and hence the magnetic anisotropy in the shell layer quite different from that of spins in the core or at the surface. Here we show that the effects of the shell layer become prominent only for very small particle sizes $D < 5$ nm.

In a recent work, Pisane *et al.* [12] reported an extension of the CSL model to account the effect of shell layer in the total effective magnetic anisotropy data of maghemite NPs. The new model considers the core-shell-surface layer (CSSL) geometry of NPs, and it has been proven successful to adequately describe the $K_{eff}$ *versus* $1/D$ data in magnetic NPs of maghemite [12] and NiO [13]. In this paper, we test the validity of the CSSL model for NPs of magnetite (Fe$_3$O$_4$) and metallic ferromagnets of Ni, Co, and Fe. The $K_{eff}$ *vs*. $D$ data used here for testing the model was taken from the published papers in literature [6, 9, 23-45], selecting data only for those NPs in which the interparticle interactions (IPI) are absent or taken into account in determining the magnitude of $K_{eff}$ vs. D. This analysis shows that the CSL model only captures the size-variation of $K_{eff}$ for larger size Ni, Co, and magnetite NPs, whereas the CSSL model adequately describes the $K_{eff}$ *versus* $1/D$ data for all particle sizes of Ni, Co, and magnetite (Fe$_3$O$_4$) NPs. The only exception appears to be the NPs of Fe for which the linear behavior of $K_{eff}$ *vs*. $1/D$ data predicted by the CSL model is valid [6]. Details of these results and discussion are presented below.

2. **Interparticle interactions and effective magnetic anisotropy:** It is important to first discuss the role of IPI and its effect on the measured blocking temperature $T_B$ which is often used to determine $K_{eff}$ (see Fig. 1(a)). The dipolar interactions between the magnetic moments of the NPs yield an additional enhancement in the magnetic anisotropy of NPs, causing a noticeable increase in $T_B$. To reduce IPI due to dipole-dipole interactions, experimentalists often use following two methods: (i) proper coating of NPs by surfactants, and (ii) dispersion/separation of NPs on a non-magnetic matrix or in suitable solvent. The strength of the IPI in blocking temperature ($T_B$) measurements can be characterized by an effective $T_0$ leading to [19]:

$$T_B = T_0 + \frac{K_{eff}V}{k_B \ln\left(\frac{f_0}{f_m}\right)} \qquad \ldots\ldots (1)$$

Here $k_B$ is the Boltzmann constant, $f_0 \sim 10^{10} - 10^{12}$ Hz is the system-dependent attempt frequency varying only weakly with temperature, $f_m$ is the experimental measurement frequency and $T_0$ is an effective temperature representing the strength of the IPI. To determine $T_0$, one can measure $T_B$ at two different measurement frequencies and evaluate the following quantities [19]:

$$\Phi = \frac{T_B(2) - T_B(1)}{T_B(1)[\log f_m(2) - \log f_m(1)]} \qquad \ldots\ldots (2)$$

$$\Phi = \Phi_o\{1 - [T_o/T_B(1)]\} \qquad \ldots\ldots (3)$$

$$\Phi_o = 2.3026/\{\ln[f_o/f_m(2)]\} \quad \ldots\ldots (4)$$

Here $T_B(1)$ and $T_B(2)$ are the blocking temperature measured at two sufficiently different frequencies $f_m(1)$ and $f_m(2)$, respectively. For no IPI ($T_0 = 0$), $\Phi = \Phi_o \sim 0.13$ and for $\Phi < 0.13$, the magnitude of IPI and $T_0$ increases with decreasing magnitude of $\Phi$ [19-22]. Since the IPI plays a crucial role in the determination of $K_{eff}$, it is important to carefully separate the contribution of IPI in the actual $K_{eff}$ value. In this work, we took care to consider only those data in which IPI were taken into account. In refs. [12, 13, 19], we have described details of the systematic evaluation of the strength of IPI using data from ac-magnetic susceptibility measurements.

3. **The Core-Shell-Surface Layer (CSSL) Model:** Figure 1(b) shows a pictorial representation of the CSSL model. The spins in core (shell) are well-ordered (partially-ordered), whereas the ordering of spins is disrupted at the surface layer due to the broken crystalline symmetry and presence of dangling bonds. The formation of shell layer is preferred because it reduces the total energy of the magnetic NPs [14]. Neutron diffraction measurements have confirmed the existence of shell layer in magnetite NPs [15]. Furthermore, by means of Monte-Carlo simulations, Kachkachi *et al.* [16] have demonstrated the formation of a shell layer of finite thickness in maghemite NPs. Due to this reason it is essential to separate the contributions of $K_{sh}$ from $K_b$ and $K_S$ in the effective magnetic anisotropy data of magnetic NPs. Here, $K_b$, $K_S$, and $K_{sh}$ are the magnetic anisotropy constants corresponding to the spins in the core, at surface layer, and in a shell of thickness $d$, respectively. According to the core-surface layer (CSL) model [6]:

$$K_{eff} = K_b + \frac{6 K_S}{D} \quad \ldots\ldots (5)$$

Here, the factor $6/D$ represents the surface/volume ratio of spherical NPs with diameter $D$. The CSSL model represents an extension of the CSL model of Eq. (5) in which we include an extra term addressing the contribution of magnetic anisotropy from spins in the spherical shell of thickness $d$ [12]:

$$K_{eff} = K_b + \frac{6 K_S}{D} + K_{sh}\left\{\left(1 - \frac{2d}{D}\right)^{-3} - 1\right\} \quad \ldots\ldots (6)$$

Here, the last term involves the ratio of the shell volume to the core volume, i.e. $\{[1-(2d/D)]^{-3} -1\}$,

and it represents the contribution of a fraction of the spins in a shell with effective anisotropy $K_{sh}$ different from $K_b$ and $K_S$. The $K_{sh}$ contribution particularly dominates the $K_b$ and $K_S$ contributions in ultra-fine magnetic NPs. For example, it was found that for maghemite NPs, the total contribution of the $K_{sh}$ term to $K_{eff}$ is about 38% for $D = 3$ nm NPs but it rapidly decreases to ~13% for $D = 4$ nm, to ~3.7% for $D = 8$ nm and to ~2% for $D = 15$ nm [12]. However, the contribution of the $K_S$ term remains significant even for $D = 20$ nm. [12] Our analysis also reveals that for smaller $d$ values, $K_{sh}$ is relatively larger due to the enhanced spin anisotropy in the shell layer, and *vice versa*. The validity of the CSSL model is limited to $D > 2d$ since only in this limit the NPs have a core. For morphologies different from a sphere, the factor 6 in Eqs. (5) and (6) should be replaced by a proper factor representing morphology of NPs.

4. **Validation of the CSSL model:** Now we test the validity of the CSSL model for four different magnetic NP systems, *viz*. Ni, Co, Fe, and $Fe_3O_4$. Fig. 2 and 3 show variation of $K_{eff}$ *vs*. $1/D$ data for these four NPs. The data were collected from available reports in the literature (references listed in the figures) where the effects due to IPI were taken into account in determining $K_{eff}$. Deviations from the linear trend of the $K_{eff}$ *vs*. $1/D$ variations predicted by the CSL model (Eq. 5) are for smaller sizes of the Ni, Co and magnetite NPs, although, Fe NPs follow the linear trend. The green line shows the best fit of the data to Eq. (5) and red line shows the best fit to Eq. (6). Although, all the data points do not exactly fall on the fitted curve, the overall trend of $K_{eff}$ *vs*. $1/D$ variation is well-captured by Eq. (6) within the experimental uncertainties.

To gain confidence in procedure used for fitting the data to Eq. (6) involving the four parameters ($K_b$, $K_S$, $K_{sh}$, and $d$), this 4-parameter problem was split into two 2-parameters problems. First, we fitted the linear part of $K_{eff}$ *vs*. $1/D$ data for larger NPs ($D > 5$ nm) using Eq. (5) thus determining the magnitudes of $K_b$ and $K_S$ from the linear fitting. Next, we used the obtained values of $K_b$ and $K_S$ values from the linear fit to determine the magnitudes of $K_{sh}$ and $d$ from fitting the data to the smaller sizes. Finally, the magnitudes of $K_b$, $K_S$, $K_{sh}$, and $d$ were fine-tuned to yield overall best fit to Eq. (6). The magnitudes of fitting parameters are given in the inset of figures. The magnitudes of the obtained magnetic anisotropy constants and the shell thickness are in excellent agreement with the reported experimental data on Ni, Co and $Fe_3O_4$ NPs. [15, 16, 40, 46, 47, 48].

5. **Conclusions:** In this paper, we have analyzed the variation of $K_{eff}$ vs. D for NPs of Ni, Fe, Co and magnetite ($Fe_3O_4$) using the available data reported in literature [6, 9, 23-45] in terms of the CSL and CSSL models discussed here. For the NPs of Ni, Co, and $Fe_3O_4$ discussed here and those of NiO [12] and maghemite [13] reported recently, the variation of $K_{eff}$ vs. $1/D$ is best described by the CSSL model. For the NPs of Fe, the CSL model appears to be quite adequate as if the Fe NPs do not have a shell. These differences for the Fe NPs might be related to how the Fe NPs were prepared or perhaps peculiar and yet un-understood physics of the Fe particles. The analysis presented here also shows that the CSSL model and hence contributions of the spins in the shell to $K_{eff}$ become important only for sizes $D <$ ~5 nm and that for larger NPs, the CSL model appears to be quite adequate to describe the linear variation of $K_{eff}$ vs. $1/D$. It is suggested that the CSSL should be tested for other magnetic NP systems when data of $K_{eff}$ vs. $D$ become available over a large enough size range with the effects of the IPI taken into account. It is hoped that the analysis presented here based on the CSSL model [12,13] will encourage the development of fundamental theoretical basis for the observed variation of $K_{eff}$ vs. D in different magnetic systems.

**Acknowledgements**: S.S. acknowledges the support of the Jefimenko Fellowship and Robert T. Bruhn Research Award at West Virginia University.

**Figures with captions:**

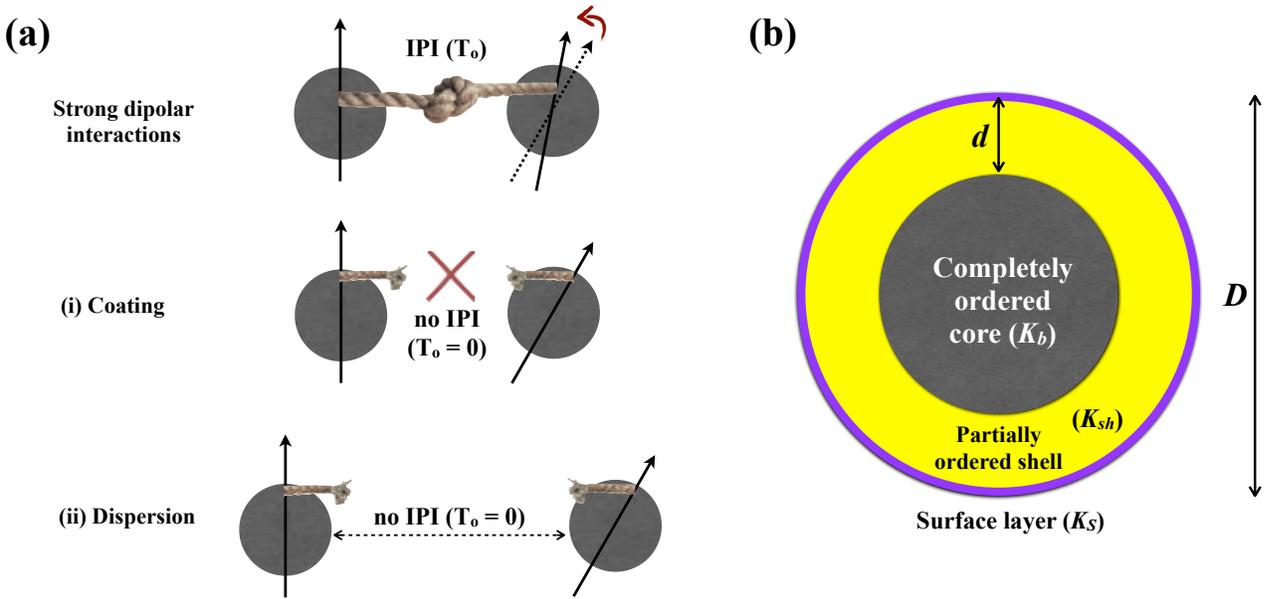

*Figure 1:* *(a) Schematic representation of the effect of IPI on the effective magnetic anistropy ($K_{eff}$) of NPs. The black arrow represents the magnetization easy axis and rope depicts the effect of IPI on $K_{eff}$. (b) Core-Shell-Surface layer geometry of a spherical magnetic NP of diameter D and shell thickness d.*

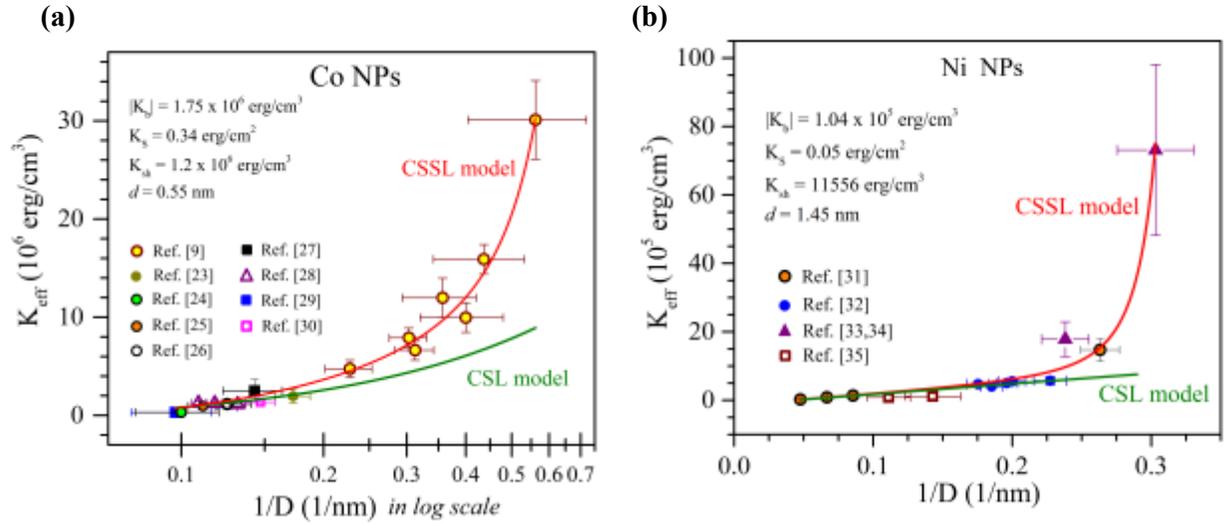

*Figure 2:* *The $K_{eff}$ vs. 1/D variation of (a) Co (x-axis is in log scale), and (b) Ni NPs. The red (green) line shows the best fit to the CSSL (CSL) model. The best fitting parameters along with the sources of the data [9, 23-35] are given in the inset.*

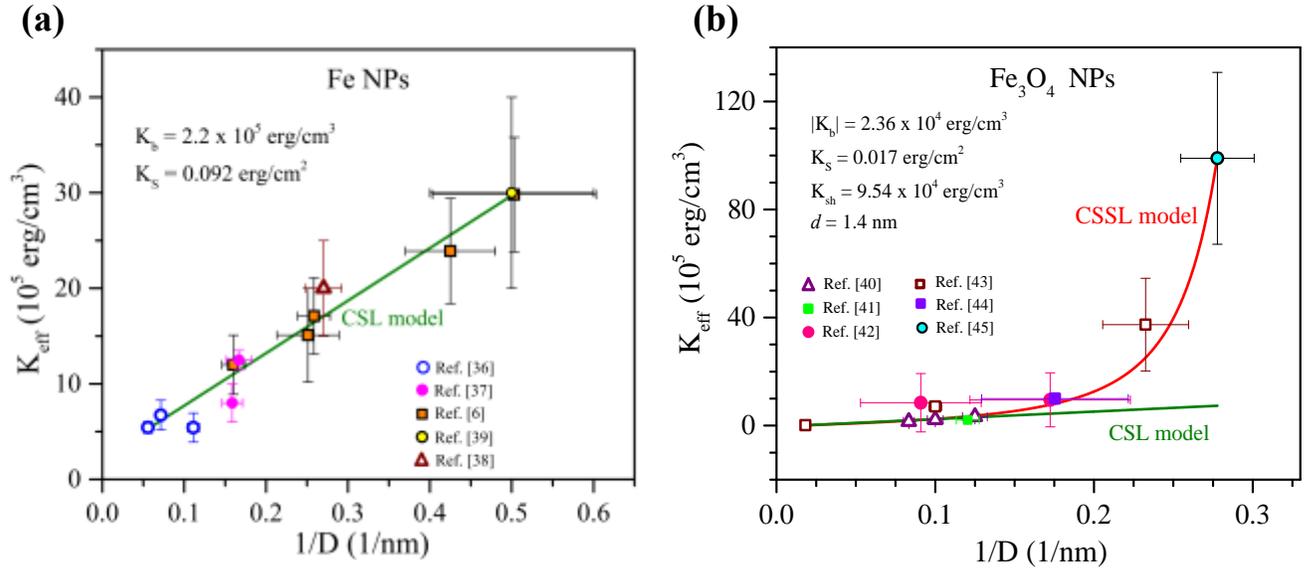

*Figure 3:* The $K_{eff}$ vs. $1/D$ variation of (a) Fe, and (b) magnetite ($Fe_3O_4$) NPs. The red (green) line shows the best fit of the data to the CSSL (CSL) model. The best fitting parameters along with the sources of the data [6, 36-45] are given in the inset.